# Biochemical analysis of human breast tissues using FT-Raman spectroscopy

*Renata Andrade Bitar[1], Herculano da Silva Martinho[1], Carlos Julio Tierra Criollo[2], Leandra Náira Zambelli Ramalho[3], Mário Mourão Netto[4], Airton Abrahão Martin[1]

[1]Laboratory of Biomedical Vibrational Spectroscopy,

Institute of Research and Development - IP&D,

University of the Valley of Paraíba - UniVap, Av. Shishima Hifumi, 2911, 12244-000,

São José dos Campos, São Paulo, Brazil

[2]Biomedical Engineering Group, Electrical Engineering Department, 31270-010, Federal University

of Minas Gerais - UFMG, Minas Gerais Brazil

[3]Department of Pathology, Ribeirão Preto Medicine College,

University of São Paulo - USP, Brazil

[4]A. C. CAMARGO HOSPITAL - Hospital do Câncer,

Rua Prof. Antonio Prudente, 211, 01509- 010, São Paulo, Brazil.

E-mail: *rabc@univap.br, carjulio@cpdee.ufmg.br, hmartinho@univap.br,

lramalho@fmrp.usp.br, amartin@univap.br

# ABSTRACT

In this work we employ the Fourier Transform Raman Spectroscopy to study normal and tumoral human breast tissues, including several subtypes of cancers. We analyzed 194 Raman spectra from breast tissues that were separated into 9 groups according to their corresponding histopathological diagnosis. The assignment of the relevant Raman bands enabled us to connect the several kinds of breast tissues (normal and pathological) to their corresponding biochemical moieties alterations and distinguish among 7 groups: normal breast, fibrocystic condition, duct carcinoma-in-situ, duct carcinoma-in-situ with necrosis, infiltrating duct carcinoma not otherwise specified, colloid infiltrating duct carcinoma and invasive lobular carcinomas. We were able to establish the biochemical basis for each spectrum, relating the observed peaks to specific biomolecules that play special role in the carcinogenesis process. This work is very useful for the premature optical diagnosis of a broad range of breast pathologies. We noticed that we were not able to differentiate inflammatory and medullary duct carcinomas from infiltrating duct carcinoma not otherwise specified.

# 1. INTRODUCTION

The breast cancer is the most common malignant tumor found in women in the Western world. Usually, the breast cancer screening involves two steps. The first one is the search for palpable lesions in the annual clinical breast examination. The second one is the X-ray mammography, in which suspicious local density changes could be detected. Whenever the tissue is particularly dense throughout, ultrasound may also be used to locate suspicious regions. If a lesion is found during the examination the tissue is submitted to biopsy that could ranges from the fine needle aspiration of single cells to the surgical removal of the entire suspicious mass by excisional biopsy. The main problem of the fine needle aspiration procedure is the high rate of false positive. Elmore *et al.*[1] have shown that over a 10 years period, during which subjects received a median of four mammograms and five clinical breast examinations, 31.7 % of all women experienced at least one false positive from either tests. Other inconveniences are related to the time expended in waiting the biopsy result as well as the biopsy procedure itself.

The breast is a large secretory gland composed of 15 to 25 autonomous and empty lobes connected to the nipple. The lobes themselves are divided into smaller units, called lobules, which are connected by ducts. Lobular and duct elements consist of single layers of epithelial and myoepithelial cells.[2] The breast undergoes many changes throughout a woman's life, both progressive due to puberty, pregnancy and menopause and cyclical due to menstruation. Hormones regulate these changes. This dynamical activity could induce a lot of opportunities for disease. Usually, breast pathology is extremely diverse, but it could be divided into two main categories: benign and malignant pathologies. Most benign breast lesions are part of a spectrum of fibrocystic changes, whereas 70 % of malignant lesions are infiltrating duct carcinomas.[2]

It is well known that the increased cell proliferation and metabolic activity in malignant tissues result in changes of the concentration and oxidation states of several biochemical species. Singer *et al.*[3] have used $^{31}$P- and $^{13}$C-nuclear magnetic resonance spectroscopy to compare the metabolite levels and fluxes through enzymes regulating phospholipid and mitochondrial metabolism of normal mammary epithelial cells with cancer cells. They found a 16-19-fold increase in the phosphocholine content in two primary breast cancer cell lines and a 27-fold increase in phosphocholine content in the metastatic breast cancer cell line compared with the normal breast epithelial cell. They also observed 30 % decrease in the ATP level, 83 % decrease in the phosphocreatine levels, 50-80 % relative reduction in the flux of pyruvate utilized for mitochondrial energy generation, along with a 2-fold increase in the NAD(+) + NADH levels in 21PT, 21NT, and 21MT-2 cells in malignant cells compared to the normal cells. This last finding was suggested as being an evidence of impaired mitochondrial metabolism in the breast carcinoma cell lines. Recently, Brull *et al.*[4] have shown that specific genetic alterations enhance the transport of choline, augment the synthesis of phosphocholine and betaine, and suppress the synthesis of choline-derived ether lipids in breast cancer cells. Through immunohistochemical and biochemical methods Yeo *et al.*[5] had shownthat there are a connection among the increased levels of altered proteoglycans and stromal desmoplasia that could explain the alterations in the extracellular matrix synthesis. Alterations in the proteoglycans appear in tumors and wound healing tissues where they present both, greater heterogeneity and longer glycosaminoglycan chains..

Similar to other kinds of cancers, the origin of breast cancer is in great extension related to multiple genetic alterations and protein dysfunctions.[6,7] In particular, the *p53* tumor suppressor gene mutation remains the most common genetic change identified in the human neoplasia.[7] Moreover, the common cellular morphological processes evolved in the malignant development as loss of differentiation, nuclear enlargement, hyperchromatism, pleomorphism and atypical mitoses could induce detectable biochemical changes, e.g., increasing nucleoproteins and nucleic acids.[8] Probing these cellular biochemical changes should provide a better understanding of the malignant tumoral process mechanism in the human tissues as well as the basis for more precise cancer diagnosis.

Raman spectroscopy is an optical technique that provides information about the molecular vibrational degrees of freedom of the investigated sample being widely used for quantitative and qualitative analytical studies in the fields of chemistry, geology, pharmacology, and solid state physics. Recently, it has emerged as a nondestructive analytical tool for the biochemical characterization of biological systems due to several advantages as sensitivity to small structural changes, non-invasive sample capability, and high spatial resolution in the case of Raman microscopy. This technique does not require wide sample preparation or pre-treatment, and making use of the FT-Raman technique employing light sources in the infrared region, the detection of weak Raman signals become easier due to fluorescence suppression. Moreover, the excitation in the near infrared, at 1064 nm, also minimizes the photo-degradation of the sample, allowing the employment of larger power densities to compensate the weak Raman signal generated by longer wavelengths.[9]

Alfano et al.[10] were the first to employ FT-Raman spectroscopy in the study of human breast tissues, using a laser at 1064 nm as excitation source. They studied 14 breast tissues being 3 normal, 4 benign, and 7 malignant, do not regarding the several subtypes of these carcinomas. They observed spectral differences between malignant, benign and normal tissues but they were unable to associate these differences to biochemical changes. Redd et al.[11] measured the Raman spectra of normal breast tissues using visible light at 406.7, 457.9, and 514.5 nm. They found differences between the peak intensities of normal and malignant tissues and attributed it to changes in the fatty acids and ß-carotene contents. Similar to Alfano et al.[10], the work of Redd et al.[11] did not concern the classification of breast carcinomas subtypes. Frank et al.[12] used excitation at 784 nm to obtain Raman spectra of normal, benign (fibro-adenoma), and malignant tissues (infiltrating duct carcinoma not otherwise specified - NOS). They noticed that the band at 1439 cm$^{-1}$ in the normal tissue shifted to 1450 cm$^{-1}$ in the infiltrating duct carcinoma NOS and they had attributed this change to the increased protein concentration in malignant samples. Using the ratio among the 1654 *and* 1439 cm$^{-1}$ band areas, they easily differentiate infiltrating duct carcinoma from normal tissue. However, they were unable to statistically differentiate among infiltrating duct carcinoma and fibro-adenoma.

Manoharan *et al.*[13] proposed the spectral classification of human breast tissues as normal, fibro-adenoma or infiltrating duct carcinoma NOS by Raman spectroscopy seeking for the identification of predominant spectral components and comparison to their histological diagnosis. Using statistical multivariate analysis based on Principal Components Analysis (PCA), Manoharan *et al.*[13] were able to correctly classify normal and pathological tissues, while benign and malignant spectra remained unclassified on that work.

Shafer *et al.*[14] have applied confocal micro-Raman spectroscopy to the study of human breast tissues. The advantage of the micro over macro-Raman spectroscopy is the relatively small laser spot size in the former. While the typical spot in macro-Raman is ~ 50-100 μm it value is ~ 5-20 μm in the micro case. In this way, Shafer *et al.*[14] were able to make a morphological/chemical image of the tissues by fitting the micro-Raman image to a linear combination of basis spectra derived from cell cytoplasm, cell nucleous, fatty acids, β-carotene, collagen, calcium hydroxiapatite, calcium oxalate dehydrate, cholesterol-like lipids, and water.

Haka *et al.*[15] studying the chemical composition with micro-Raman spectroscopy of breast duct microcalcifications, have shown that microcalcifications of calcium oxalate dihydrate occurs in benign lesions while some of those composed of calcium hydroxyapatite could be correlated to malignant lesions.

Yu *et al.*[16] investigating the micro-Raman spectra of normal and malignant breast tissues have shown the occurrence of spectral differences involving the bands of symmetric stretching modes of $PO_2^-$ group in the DNA, the symmetric stretching modes of $O=P=O$ in RNA, the bands of amide I and amide III, and the peak of the $C=O$ stretching modes in the amino acids.

Yan *et al.*[17] analyzing the Raman spectra of normal and cancerous breast cellshave shown that the intensities of the 782 and 1084 cm$^{-1}$ bands of DNA phosphate group, and 1155 and 1262 cm$^{-1}$ of deoxyribose-phosfate had decreased in the cancer cells. Moreover, the bands at 812 cm$^{-1}$ of A-type DNA and 979 and 668 cm$^{-1}$ had disappeared. The authors claim that these changes indicate that the phosphate backbone of DNA is partially destroyed in the cancer cells.

In this work we studied the Raman spectra covering the spectral region of 500 to 2100 cm$^{-1}$ of several human breast tissues to obtain a differentiation between normal tissues and 8 subtypes of breast pathologies, including fibrocystic condition, duct carcinoma-in-situ, duct carcinoma-in-situ with necrosis, infiltrating duct carcinoma-NOS, inflammatory infiltrating duct carcinoma, medullary infiltrating duct carcinoma, colloid infiltrating duct carcinoma, and invasive lobular carcinoma. Although, differentiation between normal and pathological breast tissues is well established in literature, [6, 10, 12, 14, 15, 18-27] the complete differentiation including subtypes of cancer is of special interest for clinical applications of Raman spectroscopy and, as far as we are concerned, similar study is absent in the literature.

## 2. METHODOLOGY

This research was done following ethical principles established by the Brazilian Federal Law Ministry. The patients were informed about the research and signed permission for collecting their tissue samples.

Pathological breast tissues were obtained from 30 female patients assisted in the Mastology Department of the "A. C. Camargo" Hospital, São Paulo, Brazil. Normal breast tissues were collected from 5 patients submitted to plastic surgery for breast reduction in private clinics of São José dos Campos, Brazil. The samples, soon after the surgical procedure, were identified, snap frozen and stored in liquid nitrogen (77 K) in cryogenic vials (Nalgene ®) before the FT-Raman spectra recording. For FT-Raman data collection, samples were brought to room temperature and kept moistened in 0.9 % physiological solution to preserve their structural characteristics, and placed in a windowless aluminum holder for the Raman spectra collection. Soon after, the samples were fixed in 10 % formaldehyde solution, to further histopathological analysis. We noticed that the chemical species presented in the physiological solution ($Ca^{2+}$, $Na^+$, $K^+$, $Cl^-$, water) did not have measurable Raman signal and their presence did not affect the spectral signal of the tissues.

A FT-Raman spectrometer (Bruker RFS 100/S) was used with an Nd:YAG laser at 1064 nm as excitation light source. The laser power at the sample was kept at 110 mW while the spectrometer resolution was set to 4 cm$^{-1}$. The spectra of normal and pathological breast tissues were recorded with 100 and 150 scans, respectively.

Based on the results of the histopathological diagnosis, the spectra obtained in this study were divided in nine groups: (1) normal breast; (2) fibrocystic condition; (3) duct carcinoma-in-situ; (4) duct carcinoma-in-situ with necrosis; (5) infiltrating duct carcinoma NOS; (6) inflammatory infiltrating duct carcinoma; (7) medullary infiltrating duct carcinoma; (8) colloid infiltrating duct carcinoma and (9) invasive lobular carcinoma. After subtracting the baseline, each spectrum was normalized to the maximum intensity peak (@1446 cm$^{-1}$). Afterwards, we obtained the average spectrum of each group. Typically, the average was performed over 20 spectra. We estimate the dispersion between the spectra within each group to be< 15 %. For the succeeding analysis we consider those spectra presenting a correlation coefficient better than 70 % within each group.

## 3. RESULTS AND DISCUSSION

In Fig. 1 we show the average Raman spectrum for each of the above-cited groups normalized to the 1446 cm$^{-1}$ peak. This band corresponds to C—H deformation mode of methylene group and it is nearly conformational insensitive. For this reason, it is a good standard for biological Raman spectral normalization. The spectra were vertically translated for clarity. All groups presented almost the same set of bands and the characteristic features that could differentiate normal from pathological tissues correspond to relative intensity increase/decrease of some bands or absence/appearance of some weak peaks. Several authors had pointed out this fact. [6, 10, 12, 14, 15, 18-27]

In the Table 1 we show the frequencies of the 12 main vibrational modes observed in the average Raman spectra of all groups. These frequencies were obtained after deconvoluting each spectrum into a sum of Lorentzian peaks. Comparing these data with the literature one could assign each band to a specific molecular vibration.[27-29] These assignments are presented in the Table 2 and enable one to perform a qualitative biochemical analysis of the normal and pathological breast tissues by confronting spectroscopic and histopathological data.

In Fig. 2 we show the average non-normalized Raman spectra between 1200 and 1800 cm$^{-1}$. The spectra were vertically translated for clarity and the intensity of the normal tissue was multiplied by 1/5. At first glance, the pathological tissues are immediately identified by the strong decrease in the intensity of their main bands when compared to the normal tissue. Referring to the Table 2, one could infer that the strong Raman bands at 1270, 1304, 1446, 1657, and 1747 cm$^{-1}$ can identify the large amount of lipids present in the normal breast tissue. These modes are related to the bond stretching of C—N, C=C, and C=O and bond twisting of $CH_2$, respectively. Redd *et al.*[11] identified differences in the lipids and carotene concentrations in breast tissues by Raman spectroscopy. They have shown that these bands are strongest in the normal breast than in the fibro-adenoma and carcinoma duct ones. They proposed that these spectral characteristics could promote differential diagnosis among normal tissue and benign or malignant pathologies.

Histopathologically the normal tissues analyzed in our study have shown, as main pathological features, alterations in the collagen content when compared to duct carcinoma-in-situ and infiltrating, while the others tissues presented necrosis, inflammatory cells, cysts, and DNA content variation.

In Fig. 3 and 4 we have grouped the results for normal tissue, duct carcinoma-in-situ and duct carcinoma infiltrating that present as main histopathological characteristic alterations in the collagen content. In Fig. 3 we show the average normalized Raman spectra between 800 and 1400 cm$^{-1}$. The four selected areas in Fig. 3 indicate the bands of amino acids proline, valine, glycine, and phenylalanine that characterize the primary structure of proteins. The breast pathological tissues are mainly composed of collagen [2] and proline, valine, glycine, and phenylalanine are the main collagen's amino acids.[30] It is clear the intensity increase of these bands in the tumoral tissue compared to the normal one, especially when the carcinoma becomes infiltrating. The origin of this intensity variation probably relies on the different collagen amounts present in normal and pathological tissues.[2] Moreover, it is known that the relative abundance of collagen increases in the carcinogenic process of skin,[19] lung,[31] breast,[14] and epithelial cancers in general.[27] For breast cancers in particular, due to desmoplastic reaction, also called reactive fibrosis, occurs deposition of abundant collagen as a stromal response to an invasive carcinoma. Structures relatively remote from the cancer itself may be involved such as Coopers ligaments and duct structures between the tumor and the nipple.[2] Thus, it is expected that the Raman spectra of infiltrating ductal carcinoma show more intense proline, valine, glycine, and phenylalanine collagen's bands than other tissues. This fact was just observed in fibro-adenoma,[27] infiltrating ductal,[14, 27] and invasive lobular carcinomas.[27] Considering that the infiltrating ductal carcinoma NOS is the most pathogenic of the breast cancers, the relative collagen content, as determined by Raman spectroscopy, could be used as breast cancer pathogenicity quantifier.

In Fig. 4 we show the average normalized Raman spectra between 1200 and 1800 cm$^{-1}$. The selected areas indicate the amide I and amide III bands corresponding to vibrational modes of the peptide bonds of the secondary structure of proteins. These peaks also became more intense in the malignant tissues, especially in the infiltrating carcinoma. The amide I band corresponds to vibration of C, O, and H atoms in the –CONH– group.[28] The amide III band involves the motion of C–radical, C–N, and N–H groups.[28] Mahadevan-Jansen *et al.*[29] also observed these bands by studying the role of the proteins in the benign and malignant breast tissues. Beyond the spectral differences between benign and malignant tissues our results show that there are measurable differences in the secondary structure of proteins of breast tissues that could differentiate subtypes of malignant lesions, as ductal carcinoma-in-situ and infiltrating ductal carcinoma, and jointly with collagen content could be used to quantify the degree of pathogenicity of breast cancer.

In Fig. 5 we compare the duct carcinoma-in-situ with and without necrosis analyzing the spectral differences of the necrotic elements. In this figure we show the average normalized Raman spectra of the duct carcinoma-in-situ with and without necrosis in the spectral range of 800 to 1800 cm$^{-1}$. In the stippled area we indicate the low intensity bands that shown a relative intensity increase in the tissue with necrosis. As shown in the Table 2, the peaks in the region of 1304-1310, ~ 1446, and 1657-1660 cm$^{-1}$ are vibrational modes of adenine, cytosine, collagen, lipids, carbohydrates, proteins, and pentoses, respectively. These bands are very similar to those seen by Shafer *et al.*[14] in the cholesterol sample simulating necrosis. The intensity differences could be related to the presence of necrosis and lymphocytes in the tissue with necrosis. The lymphocytes appear due to the inflammatory process in this tissue sample. Thus, this spectral region could be useful to diagnose the presence of necrosis in breast cancer tissues.

In Fig. 6 we compare fibrocystic condition to infiltrating ductal carcinoma NOS and colloid infiltrating ductal carcinoma in order to analyze the cystic spectral features. In this figure we show the average normalized Raman spectra of the fibrocystic, infiltrating duct and colloid carcinoma tissues in the spectral range of 800 to 1200 cm$^{-1}$. The selected areas indicate the main bands that suffer alterations. The intensity variation of the bands in the interval of 800 to 1000 cm$^{-1}$ is related to different collagen contend, as just commented in Fig. 3. Carbohydrate bands that are related to the amorphous substance of the cystic content dominate the region between 1020 and 1140 cm$^{-1}$. The vibrational bands at 1065 and 1085 cm$^{-1}$ are related to the —CH—OH bond and C—O stretching coupled to the C—O group of the carbohydrates. The bands at 1050, 1065 and 1150 cm$^{-1}$ of infiltrating duct carcinoma, fibrocystic and colloid infiltrating duct carcinoma present an increasing intensity. This fact could be used to classify these three subtypes of carcinomas, and it refers to the different colloid content present in the samples.

In Fig. 7 and 8 we compare the normal tissue, carcinoma ductal-in-situ, infiltrating ductal carcinoma and invasive lobular carcinoma relating the invasiveness potential to the cysteine content and C≡C band. Fig. 7 shows the average normalized Raman spectra in the region of 500 to 580 cm$^{-1}$. We found that there are special features in this region that could be of interest for the clinical diagnosis between normal, "in situ" and infiltrating tissues. The most intense peak, at 538 cm$^{-1}$, is related to disulphide bridges in cysteine and it presents a strong intensity increase when compared to the normal tissue. The invasive tumors showed more intense peaks than carcinoma-in-situ. Thomssen *et al.*[32] have shown that the lysosomal cysteine proteases cathepsin B and cathepsin L have been implicated in tumor spread and metastasis, underlining the significance of tumor-associated proteolysis for invasion and metastasis. In this way the observed intensity variation of the Raman band at 538 cm$^{-1}$ in normal tissue, "in situ" and infiltrating carcinomas could be well understood when keeping in mind these changes in cysteine content occurring in tumoral process. For this reason, this spectral region could be used to perform initial diagnosis and real-time differentiation between "in situ" and infiltrating lesions. Unfortunately from our data we were not able to differentiate infiltrating ductal from infiltrating lobular carcinomas.

Fig. 8 shows the average normalized Raman spectra in the region between 2000 and 2100 cm$^{-1}$ for the same set of tissues of Fig. 7. One could observe that the peaks at 2028 and 2084 cm$^{-1}$ for ductal carcinoma-in-situ present a 5-fold intensity increase compared to normal tissue. For infiltrate ductal and invasive lobular carcinomas this intensity increase is nearly 10 times. Thus, the spectral differences in these two bands could be used to distinguish among tissues with these malignancies and normal tissues. These two peaks are related to C≡C bonds [28] that could be present in unsaturated fatty acids, lipids and steroids. Unfortunately, we could not identify specific biomolecules responsible for this vibration in breast tissues. The low intensity peak at 2062 cm$^{-1}$ for ductal carcinoma-in-situ, invasive lobular carcinoma and normal tissue is near to the noise level and we will not analyze it. We noticed that we are not able to differentiate inflammatory and medullary duct carcinomas from infiltrating duct carcinoma NOS.

# 4. CONCLUSION

In summary, in this study we analyzed the Raman spectra of normal and tumoral breast tissues, including several subtypes of cancers, searching for specific spectral features that could differentiate normal and pathological tissues. The collected samples were histopathologically classified into 9 groups according to their morphological features. Through the qualitative analysis of the Raman spectra and assignment of the relevant bands from the literature, it was possible to built spectral models and differentiates among 7 groups normal breast, fibrocystic condition, duct carcinoma-in-situ, duct carcinoma-in-situ with necrosis, infiltrating duct carcinoma NOS, colloid infiltrating duct carcinoma and invasive lobular carcinoma. These differences were established through the comparative study between the spectral differences and the histopathological diagnosis. Furthermore, we were able to establish the biochemical basis for each spectrum by relating the observed peaks to specific biomolecules that have special role in the carcinogenesis process. We noticed that we were not able to differentiate inflammatory and medullary duct carcinomas from infiltrating duct carcinoma NOS. Our results were summarized in Fig. 9. This work is very useful for the precocious optical diagnosis of a broad range of breast pathologies and as far as we are concerned, similar work had not been made in the literature.

## ACKNOWLEDGMENTS

Authors wish to thank Dr. Ronaldo Roesler for providing breasts samples, and A.A.M. thanks CNPq (302393/2003-0) and FAPESP (2001/14384-8) for providing financial support.

Tables

Table 1- Vibrational mode frequencies (cm$^{-1}$) observed by Raman spectroscopy in the human breast tissues groups (1) normal; (2) fibrocystic; (3) duct carcinoma-in-situ; (4) duct carcinoma-in-situ with necrosis; (5) infiltrating duct carcinoma NOS; (6) inflammatory infiltrating duct carcinoma; (7) medullary infiltrating duct carcinoma; (8) colloid infiltrating duct carcinoma, and (9) invasive lobular carcinoma.

| Peaks | (1) | (2) | (3) | (4) | (5) | (6) | (7) | (8) | (9) |
|---|---|---|---|---|---|---|---|---|---|
| 1 | - | 538.9 | 538.95 | 541.35 | 538.92 | 538.68 | 538.72 | 538.76 | 538.64 |
| 2 | 858.75 | 861.7 | 858.4 | 860.75 | 854.08 | 853.34 | 854.57 | 869.38 | 853.51 |
| 3 | - | 935.6 | - | 942.87 | 964.83 | 970.27 | 962.09 | 976.14 | 971.26 |
| 4 | 1005 | 1005.3 | - | 1005.1 | 1005.4 | 1004.7 | 1004.8 | 1007.1 | 1005.7 |
| 5 | 1080.6 | 1091.4 | 1080.7 | 1076.2 | 1082.4 | 1084.4 | 1095.6 | 1091.6 | 1083.5 |
| 6 | 1266.6 | 1261.2 | 1264.1 | 1264.2 | 1262.5 | 1261.1 | 1274.5 | 1277.2 | 1266.9 |
| 7 | 1304 | 1305.4 | 1304.8 | 1310.6 | 1305.2 | 1303.8 | 1303.3 | 1305.4 | 1303.6 |
| 8 | 1446.4 | 1449.4 | 1447.5 | 1450.6 | 1448.6 | 1448 | 1450.2 | 1449.6 | 1447.8 |
| 9 | 1657.1 | 1657.7 | 1657.8 | 1643.0 | 1659.5 | 1659.1 | 1659.7 | 1657.1 | 1657.8 |
| 10 | 1747.4 | 1747.6 | 1746.4 | - | 1748.1 | 1746.4 | 1743.5 | 1748 | 1747.8 |
| 11 | - | - | 2028.9 | - | 2028.9 | 2029 | 2029.3 | 2028.5 | 2029 |
| 12 | - | - | 2084.2 | - | 2084.2 | 2083.7 | 2083.6 | 2079.1 | 2084.3 |

Table 2 - Assignment of the main regions observed by Raman spectroscopy in normal and pathological human breast tissues. δ and ν correspond to stretching and twisting vibrational modes of the corresponding chemical bond.[27-29]

| Peaks position (cm$^{-1}$) | Vibrational Mode | Major Assignments |
|---|---|---|
| ~ 538 | $S-S$ | disulphide bridges in cysteine |
| 853-869 | $\nu(C-C)$, ring breathing, $\nu(O-P-O)$ | proline, tyrosine, DNA |
| 935-975 | $\nu(C-C)$, α-helix | proline, valine, protein conformation, glycogen |
| ~ 1005 | symmetric ring breathing mode | phenylalaline |
| 1080-1095 | $\nu(C-C)$ or $\nu(C-O)$, $\nu(C-C)$ or $\nu(PO_2)$, $\nu(C-N)$, $\nu(O-P-O)$ | lipids, nucleic acids, proteins, carbohydrates |
| 1260-1275 | $\nu(C-N)$ of amide III, $\nu(=C-H)$ | proteins (α-helix), lipids |
| 1304-1310 | $\delta(CH_2)$, $\delta(CH_3CH_2)$ | adenine, cytosine, collagen, lipids |
| ~ 1446 | $\delta(CH_2)$ | lipids, carbohydrates, proteins and pentose |
| 1657-1660 | $\nu(C=O)$ of amide I, $\nu(C=C)$ | proteins (α-helix), lipids |
| ~ 1746 | $\nu(C=O)$ | lipids |
| ~ 2028, 2062, 2084 | $\nu(C \equiv C)$ | lipids, fatty acids or hormones |

Figure Captions

Figure 1 – Normalized average Raman spectrum of (1) normal; (2) fibrocystic condition; (3) duct carcinoma-in-situ; (4) duct carcinoma-in-situ with necrosis; (5) infiltrating duct carcinoma; (6) infiltrating duct carcinoma inflammatory; (7) infiltrating duct carcinoma medullar; (8) infiltrating duct carcinoma colloid and (9) infiltrating lobular carcinoma groups. The spectra were vertically translated for clarity.

Figure 2 – Non-normalized average Raman spectrum of (1) normal; (2) fibrocystic condition; (3) duct carcinoma-in-situ; (4) duct carcinoma-in-situ with necrosis; (5) infiltrating duct carcinoma NOS; (6) inflammatory infiltrating duct carcinoma; (7) medullary infiltrating duct carcinoma; (8) colloid infiltrating duct carcinoma, and (9) invasive lobular carcinoma in the spectral region of 1200 to 1800 $cm^{-1}$. The spectra were vertically translated for clarity and the intensity of normal tissue (1) was multiplied by 1/5.

Figure 3 - Normalized average Raman spectra between 800 and 1400 $cm^{-1}$ for normal tissue and in-situ/infiltrating duct carcinomas.

Figure 4 - Normalized average Raman spectra between 1200 and 1800 $cm^{-1}$ for normal tissue and in-situ/infiltrating duct carcinomas. The selected areas indicate the amide I and amide III bands corresponding to the peptidic bonds in the secondary structure of proteins.

Figure 5. Normalized average Raman spectra of the "in situ" duct with and without necrosis tissues in the spectral range of 800 to 1800 $cm^{-1}$. In the stippled area are indicated the low intensity bands that are very similar to those seen by Shafer *et al.*[14] in the cholesterol sample simulating necrosis.

Figure 6. Normalized average Raman spectra of the fibrocystic condition, infiltrating duct carcinoma NOS and colloid type in the spectral range of 800 to 1200 $cm^{-1}$.

Figure 7. Normalized average Raman spectra in the region of 500 to 580 cm$^{-1}$. The most intense peak at 538 cm$^{-1}$ represent the amino acid cysteine.

Figure 8. Normalized average Raman spectra in the region between 2000 and 2100 cm$^{-1}$. The peaks at 2028 and 2083 cm$^{-1}$ are related to proteins and nucleic acids.

Figure 9. Summary of the several breast cancers classes studied in this work and their corresponding main Raman spectral differences. At left we present the main biomolecular vibration enabling one to discriminate the specific kind of tissue indicated in the right.

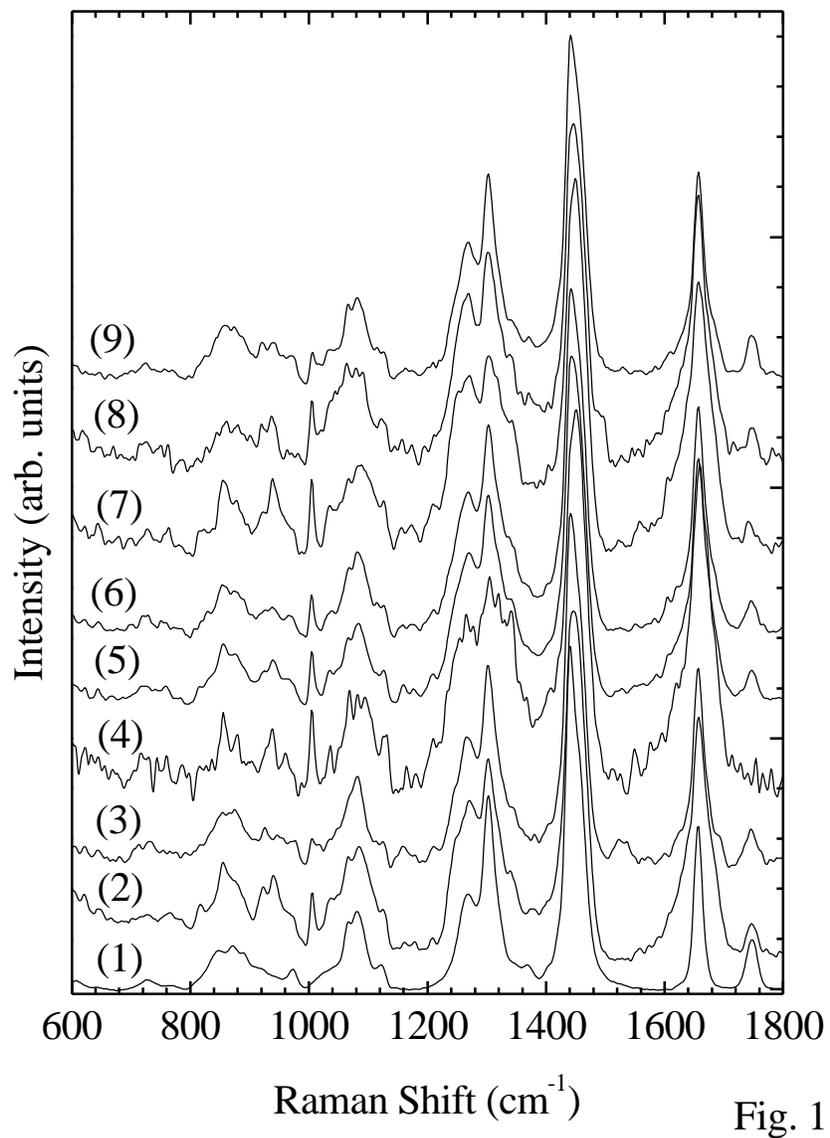

Fig. 1

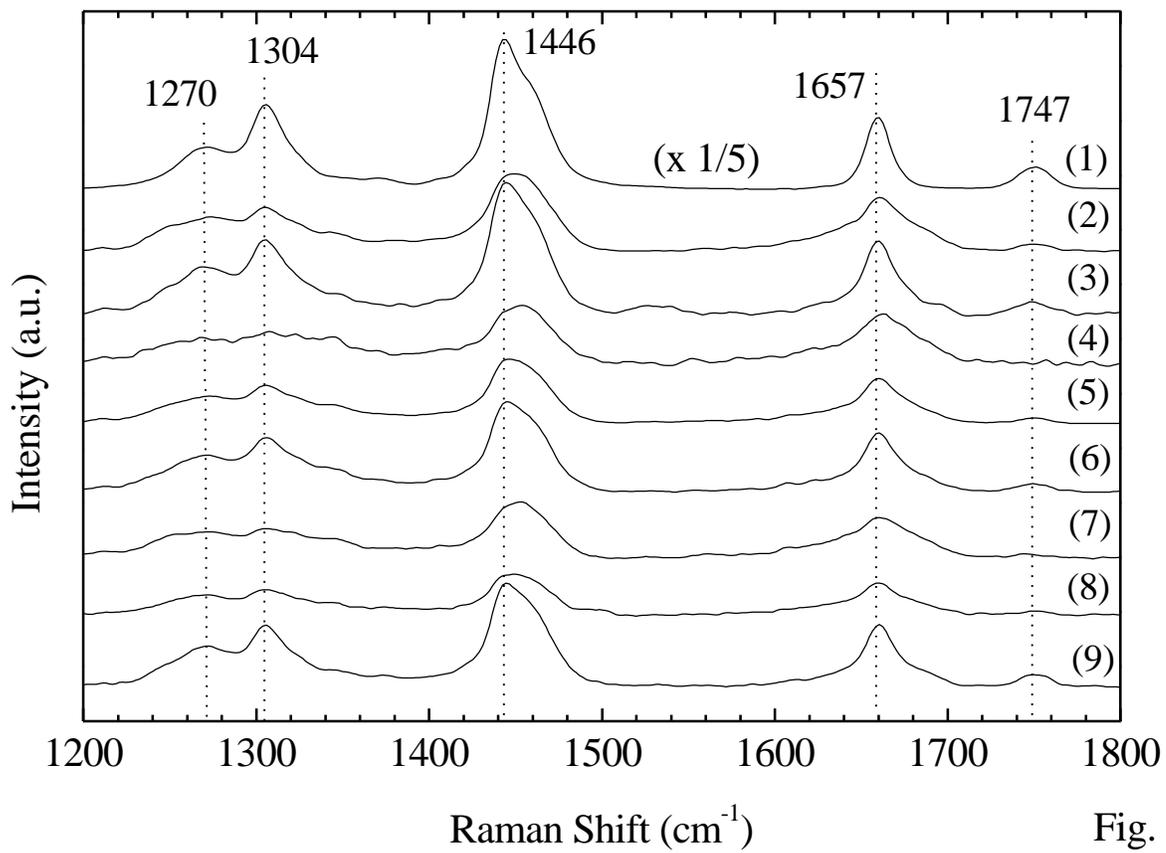

Fig. 2

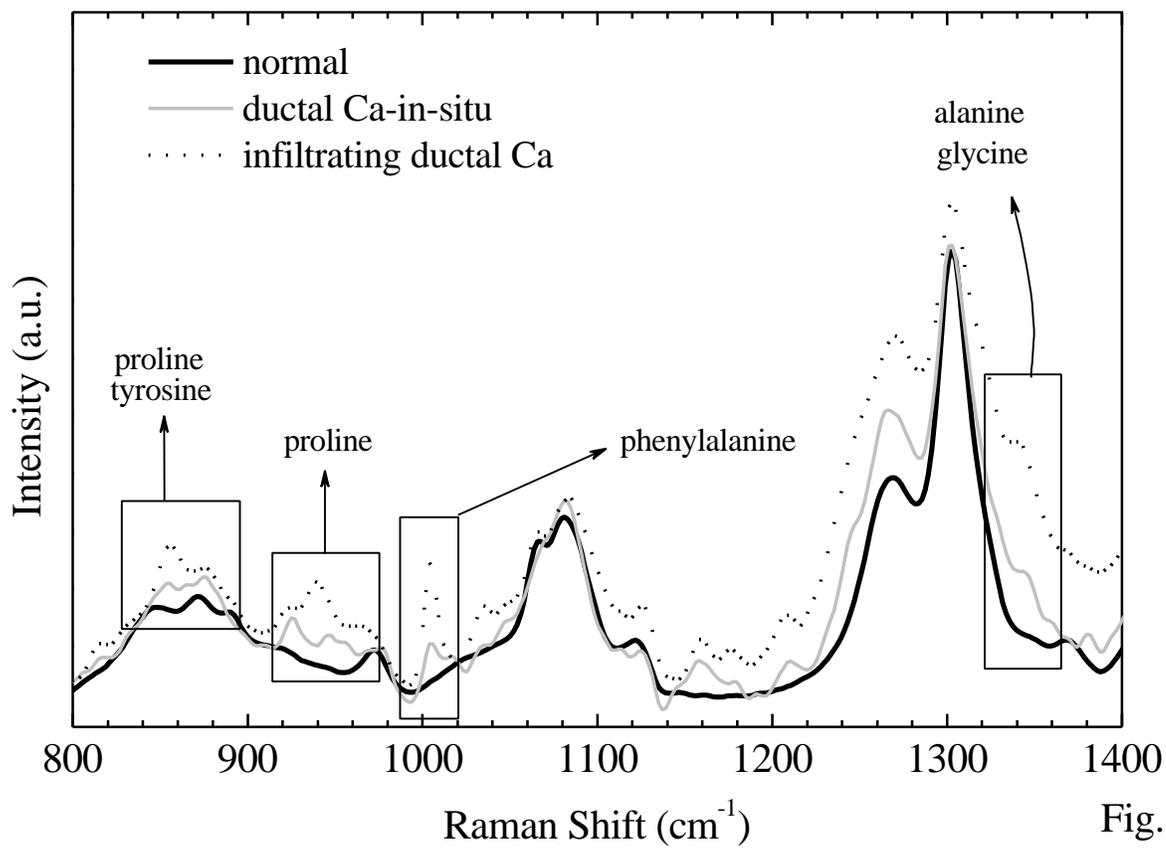

Fig. 3

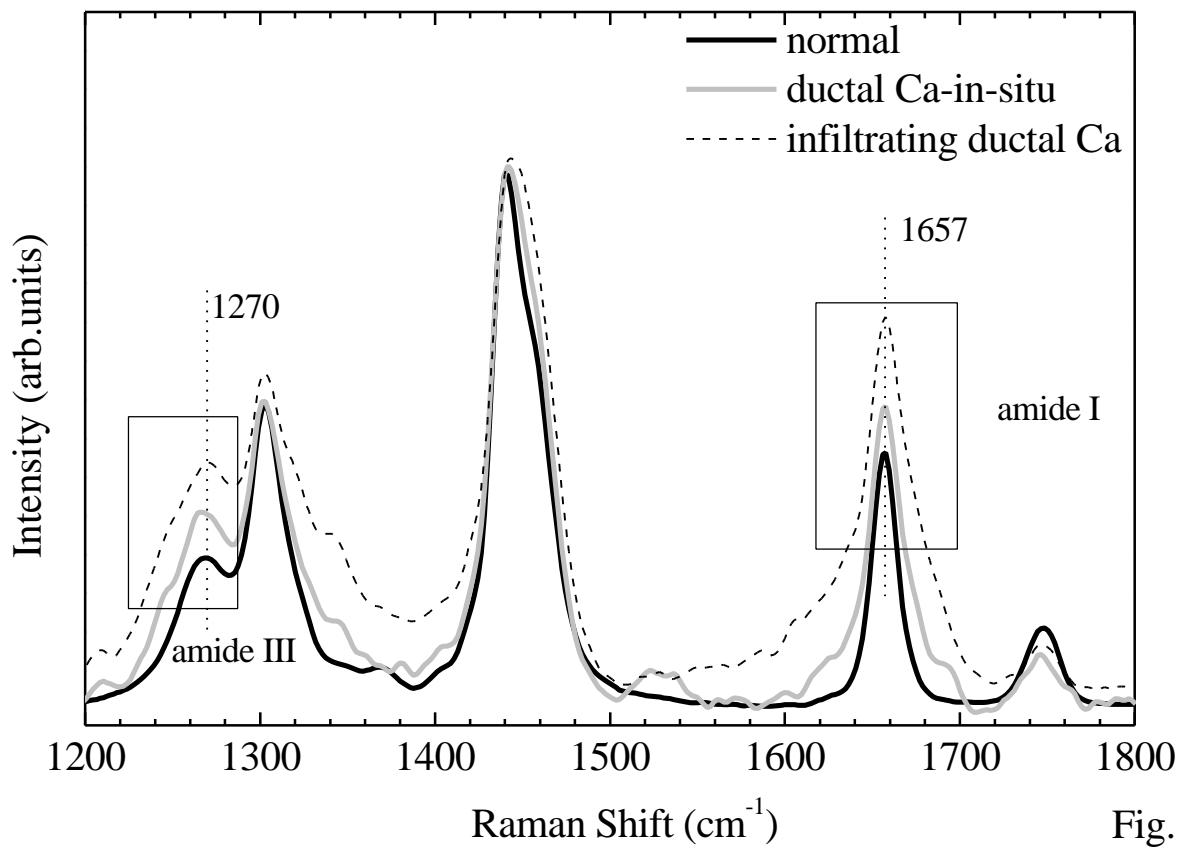

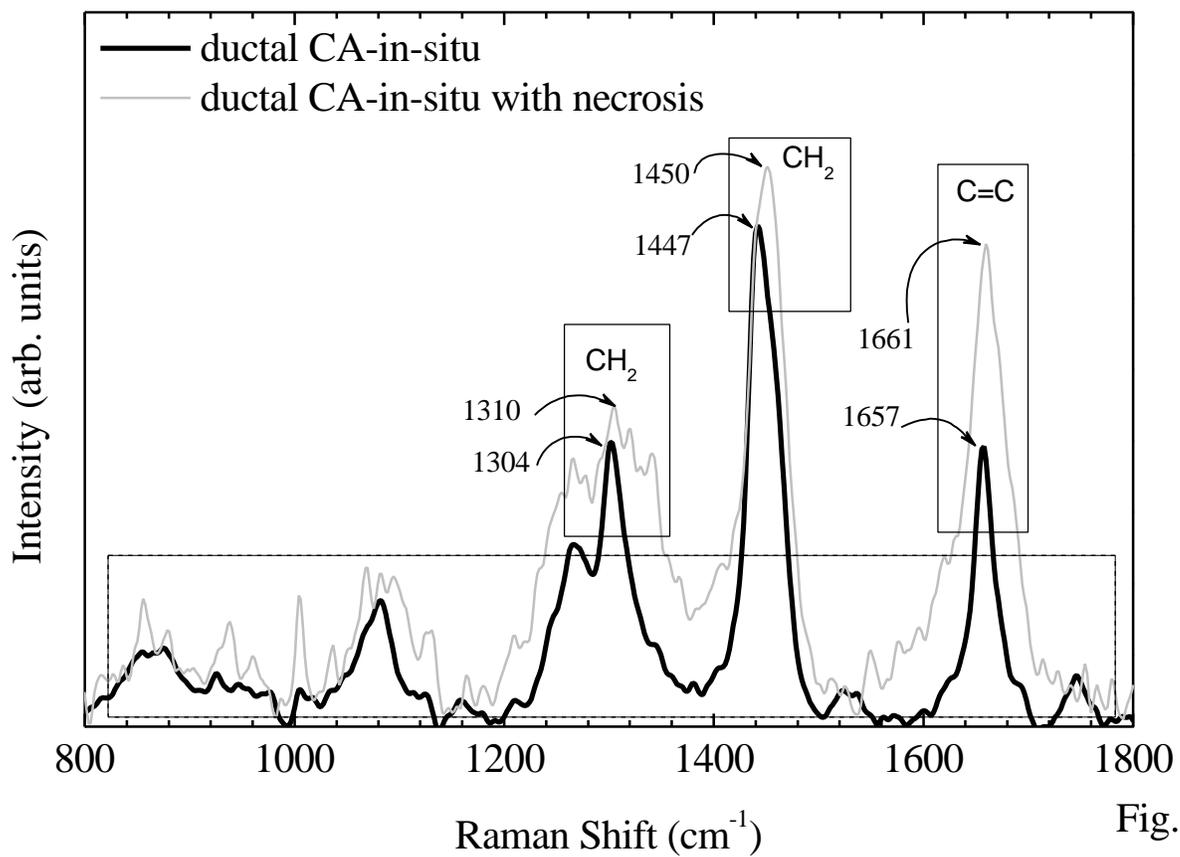

Fig.5

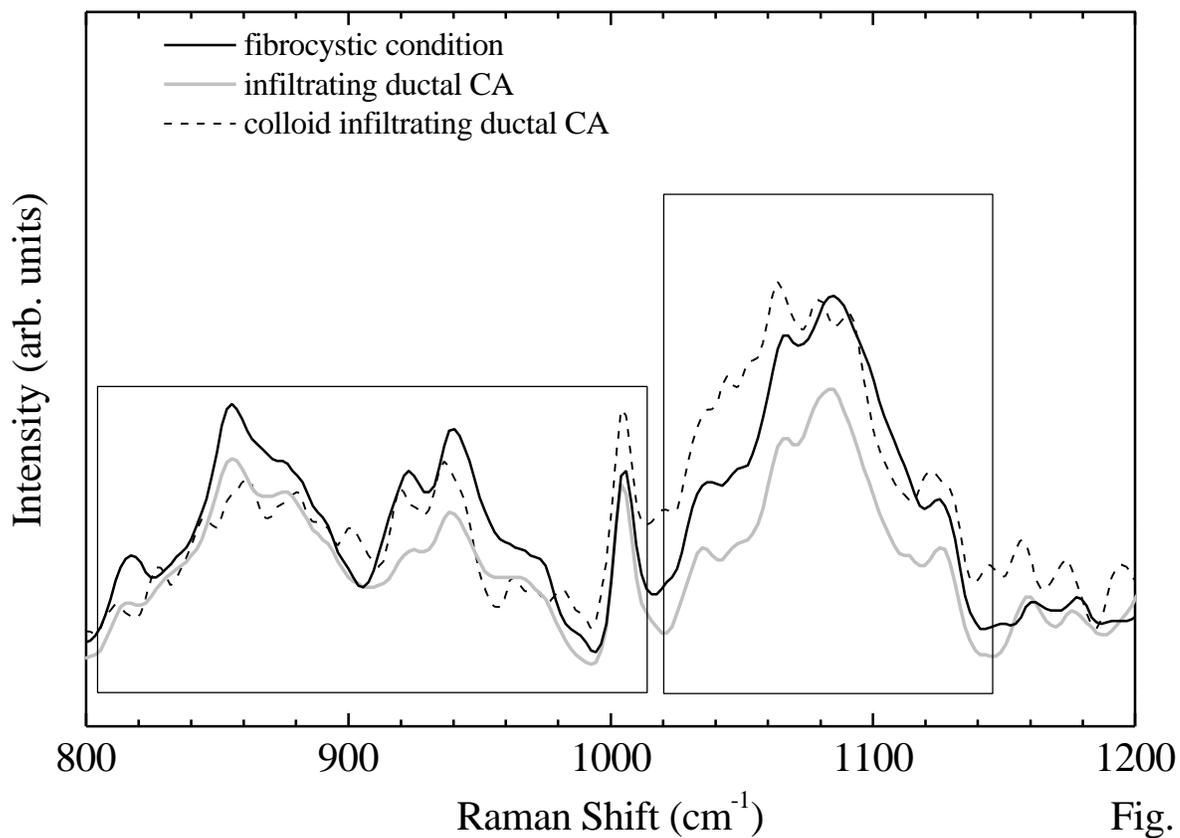

Fig. 6

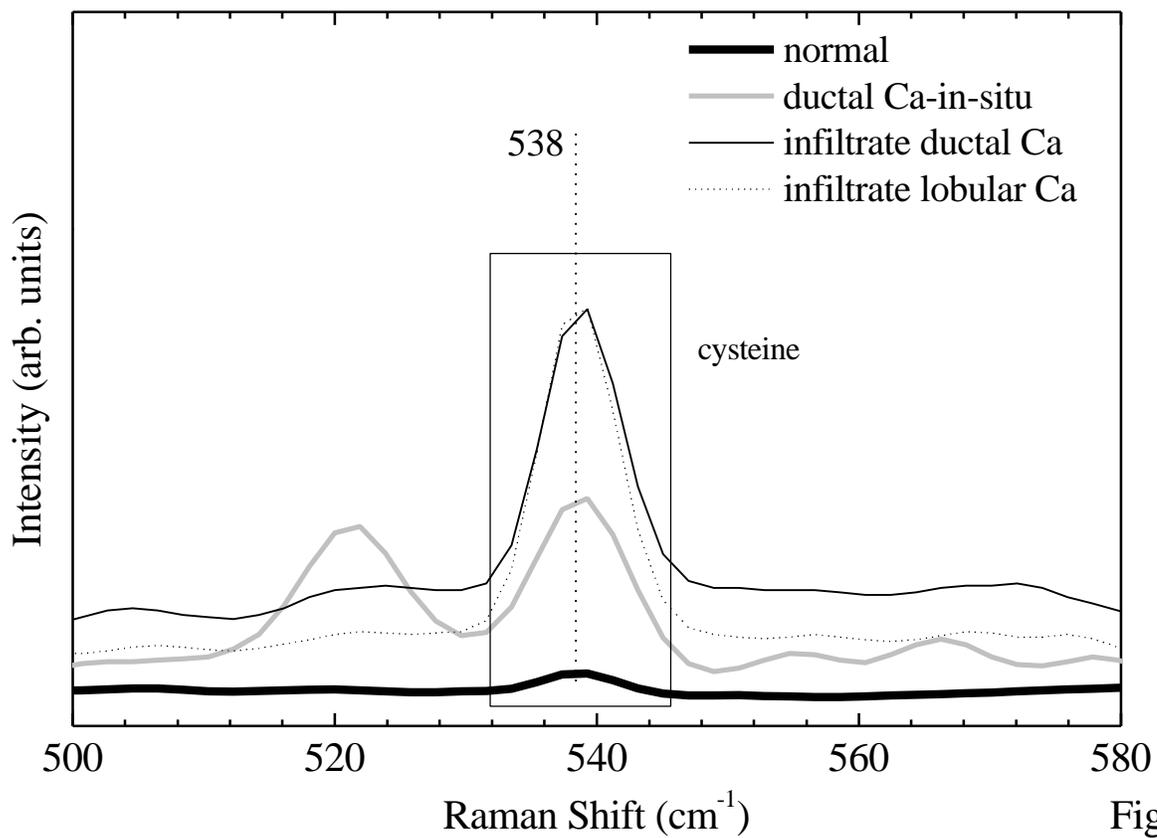

Fig.7

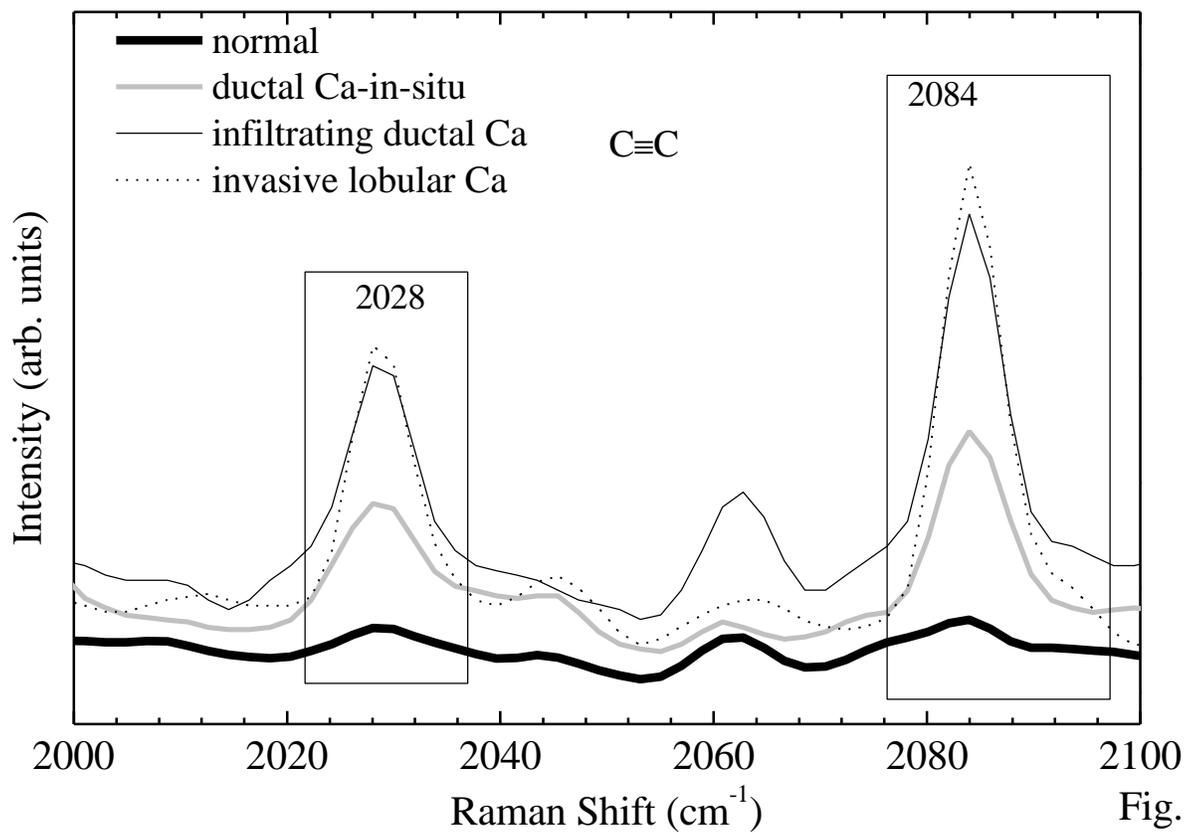

Fig. 8

**Primary structure of proteins**
**(800 – 1400 cm$^{-1}$)**
- normal breast tissue
- in situ duct CA
- infiltrating duct CA-NOS

**Carbohydrates**
**(800 – 1000 cm$^{-1}$ and 1000-1150 cm$^{-1}$)**
- fibrocystic
- infiltrating duct CA-NOS
- colloid infiltrating duct CA

**Cysteine/ C=C**
**(540 cm$^{-1}$/ 2028 and 2084 cm$^{-1}$)**
- normal
- duct CA-in-situ
- infiltrating duct CA-NOS
- invasive lobular CA

**Lipids, Cholesterol**
**(1200 – 1600 cm$^{-1}$)**
- duct CA-in-situ
- duct CA-in-situ with necrosis

**Amide I/ Amide III**
**(1200 – 1400 cm$^{-1}$/ 1600 – 1700 cm$^{-1}$)**
- in-situ duct CA
- normal breast tissue
- infiltrating duct CA-NOS

Fig. 9